\documentclass{aastex62}

\usepackage{epstopdf}
\usepackage{amsmath}
\usepackage{natbib}
\usepackage{hyperref}

\shorttitle{ Neutral Chlorine Oscillator Strengths}
\begin{document}
\title{Lifetimes and Oscillator Strengths for Ultraviolet Transitions in Neutral Chlorine}
\correspondingauthor{R.B. Alkhayat} \\
\email{ rabee.alkhayat@rockets.utoledo.edu, richard.irving@utoledo.edu, steven.federman@utoledo.edu, david.ellis@utoledo.edu, song.cheng@utoledo.edu.} 

\author{R.B. Alkhayat}
\affiliation{Department of Physics and Astronomy, University of Toledo, Toledo, OH 43606}
\author{R.E. Irving}
\author[0000-0002-8433-9663]{S.R. Federman}
\author{D.G. Ellis}
\author{S. Cheng}

\begin{abstract}

We present lifetime measurements using beam-foil techniques for radiative transitions from the 3$p^4$($^1S$)4$s$ $^2S_{1/2}$, 3$p^4$($^3P$)5$s$ $^2P_{1/2,3/2}$, and 3$p^4$($^3P$)3$d$ $^2F_{5/2}$ levels in Cl~{\footnotesize I} and the corresponding results of the oscillator strengths for transitions at 1004.68, 1079.88, 1090.73, and 1094.77 \AA, respectively. We compare our experimental results with available theoretical calculations and astronomical observations in an effort to resolve discrepancies among them.  

\end{abstract}
%
%
%
%
%
%

\section{Introduction} 
Neutral  chlorine plays an important role in understanding the structure of diffuse interstellar clouds. A small fraction of the chlorine in diffuse clouds is in the form of HCl, while most of the chlorine converts to HCl in dense clouds because of the reaction of Cl with H$_3^+$ \citep{dal74}. Knowledge of various ionization states of chlorine \citep{jur78,moo12} allows a better understanding of the processes affecting diffuse cloud chemistry \citep{jur74,neu09}. Cl~{\footnotesize I} is an effective way of tracing the molecular fraction of H$_2$ \citep{son02}. Furthermore, Cl~{\footnotesize I} has over $50$ transitions between 920 and 1390 \AA\ \citep{mor03}, but few of the most prominent ones have reported experimental lifetimes.

The oscillator strengths of the Cl~{\footnotesize I} lines are difficult to compute from ab initio theory because there is strong configuration interaction and spin-orbit mixing, especially between the levels of the $3p^4$ 3$d$ and $3p^4$ 5$s$ configurations. For this reason, LS-coupling or any other coupling scheme seems not to describe the states very well. This creates a situation where the calculation of the oscillator strengths is quite challenging and leads to confusion in state designations.

Few theoretical and experimental data of transition probabilities and lifetimes are available for Cl~{\footnotesize I}. \cite{ojh90} performed a configuration interaction calculation using CIV3 on Cl~{\footnotesize I} to compute a large number of transitions. However, they recommend caution when using their results due to the strong mixing between $3p^4$ $(^3P)$ 3$d$ and  $3p^4$ $(^3P)$ 5$s$.  \cite{bie94} reported oscillator strengths for Cl~{\footnotesize I} by considering the most important interaction and relativistic effects. They suggested additional elaborate work, especially for weaker transitions. The early theoretical work of \cite{ojh90} and \cite{bie94} shows significant  discrepancies in oscillator strengths  with astronomical and experimental results \citep{fed86,sch93}, particularly  the lines $\lambda\lambda$1088, 1097 for the  reasons mentioned earlier. \cite{sin06} calculated  oscillator strengths using configuration interaction wavefunctions, but their results show  differences in the length and velocity forms. Astronomical observations were described by \cite{son06} for the Cl~{\footnotesize I} lines at 1004.68, 1079.88, 1090.73, and 1094.77 \AA. All quoted wavelengths are given in vacuum.  [NIST wavelengths as well as designations for energy levels \citep{kra08} are used in this paper, except for the changes in designation suggested by \cite{oli07} noted below.] However, their resulting oscillator strengths are different from theoretical predictions \citep{bie94} by factors~3.1, 1.2, 2.4, and 0.42, respectively. \cite{oli07,oli08,oli13} published a set of large-scale calculations for neutral chlorine in an attempt to understand the causes for the continuing discrepancies between theory and observation or experiment; they point to the strong  mixing between the $J=5/2$ levels of the $3p^4$ 3$d$ and $3p^4$ 5$s$ configurations.

The main plan of this paper is to provide experimental results for lifetimes and oscillator strengths through beam-foil spectroscopic techniques and to resolve the existing conflicts between  theory versus observations and experiment. No previous experimental values have been reported to estimate the oscillator strengths for the lines of interest to \cite{son06}.

\section{Experimental Details} 

The experimental lifetime measurements were executed at the Toledo Heavy Ion Accelerator (THIA). More details about the experimental techniques and the facility were described in earlier published papers \citep[e.g.,][]{fed92,haa93, sch00}. Chlorine ions were produced by heating sodium chloride in a high-temperature oven at the THIA accelerator using a Danfysik Model 911A Universal Ion Source and then selected by a bending magnet. The ion beam was  accelerated to final kinetic energies of 130 or 170 keV. 
Chlorine ions were steered toward carbon foils with thicknesses ranging from 2.2 to 2.4 $\mu$g cm$^{-2}$ by an electrostatic switchyard. Typical beam currents were set to be around 90  to 100 nA to reduce foil breakage. An Acton 1 m normal-incidence vacuum ultraviolet monochromator with a 2400 line mm$^{-1}$ grating blazed at 800 \AA\  was used for the transitions for the various charges and excited states produced when ions passed through a carbon foil. A Galileo Channeltron detected the emitted radiation at the exit slit of the monochromator. The spectrometer has a resolution of approximately 0.2 \AA.

In this study, the lifetimes and corresponding oscillator strengths of lines at 1004.68, 1079.88, 1090.73, and 1094.77 \AA\ were obtained. Because the lines under investigation are relatively weak, only forward decay curves were acquired by moving the carbon foils upstream away from the monochromator slit. The error bars associated with the lifetimes take into account both experimental and statistical uncertainties. The fluctuation in signals present in spectra and decay curves is less than 20\%. All decay curves and spectral data were normalized  using an optical monitor signal from a photomultiplier tube. Possible systematic effects such as nuclear scattering, foil thickening, and beam divergence were studied at two different energies \citep[see][]{fed92} in the forward direction, and the spread among the results for a transition yielded the value for the systematic uncertainty (typically 5$\%$).  The statistical uncertainties (about 6$\%$) were inferred from the random noise in a specific measurement.  The systematic errors were added in quadrature to the statistical errors to determine uncertainties in the lifetime measurements.

\begin{figure}[ht]
\centering
\includegraphics[width=0.7\textwidth]{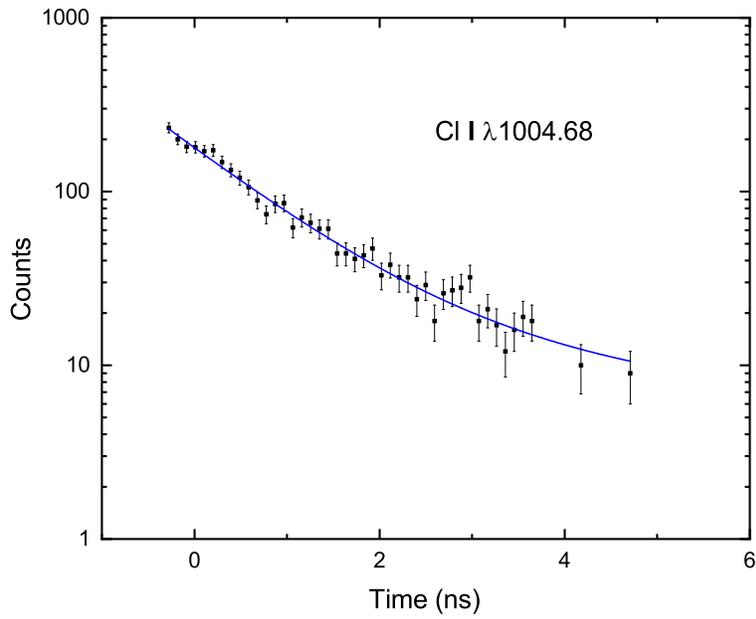}
\caption{Decay curve for the line at 1004.68 \AA\ for a beam energy of 170 keV. The post-foil velocity of 0.941 mm ns$^{-1}$ was used to convert the foil position into a time. The foil was moved relative to the monochromator entrance slit in increments of 0.1 mm until it was displaced 3.7 mm; then the increments were increased to 0.5 mm. The curvature beyond about 3 ns in the fitted decay curve (blue line) arises from the presence of decays in Cl~{\footnotesize III} as noted in the text.}
\label{fig:f1}
\end{figure}

Using the line at 1004.68 \AA, the decay curve for the 3$s^2$3$p^4 4s$ $^2S_{1/2}$ level was affected by a line at 1005.28 \AA\ due to Cl~{\footnotesize III}. To account for blending from this line, we took decay curves at two settings, namely 1004.53 \AA\ and 1005.00 \AA.  Analysis and comparison of these two decay curves confirmed the influence of the blending with the Cl~{\footnotesize III} line and gave results consistent with its previously measured meanlife of 7.87 ns \citep{sch05}.  With this information, we performed a two-exponential fit, as shown in Figure~\ref{fig:f1}, to extract the desired meanlife for the 3$s^2$3$p^4$ 4$s$  $^2S_{1/2}$ level.

\begin{figure}[ht]
\centering
\includegraphics[width=0.7\textwidth]{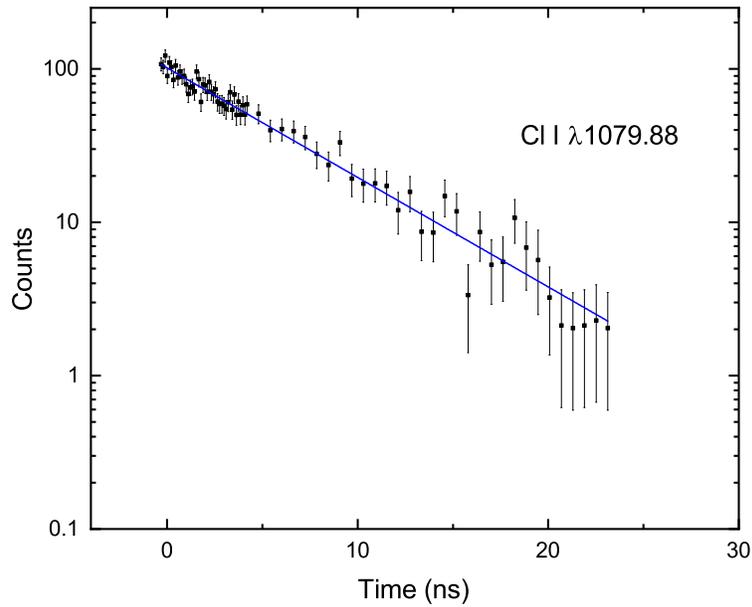}
\caption{Decay curve for the line at 1079.88 \AA\ for a beam energy of 130 keV. The post-foil velocity of 0.818 mm ns$^{-1}$ was used to convert the foil position into a time.}
\label{fig:f2}
\end{figure}

A single exponential fit was adequate to obtain the lifetime from decay of the 3$s^2$3$p^4$5$s$ $^2P_{1/2}$ level seen via $\lambda$1079.88. To the precision of our measurements, this transition was not affected by  cascades, nor by the presence of the nearby strong line of Cl~{\footnotesize II} at 1079.08 \AA\ whose lifetime is 9.14 ns \citep{sch05}. The fitted decay curve appears in Figure~\ref{fig:f2}.

\begin{figure}[ht]
\centering
\includegraphics[width=0.7\textwidth]{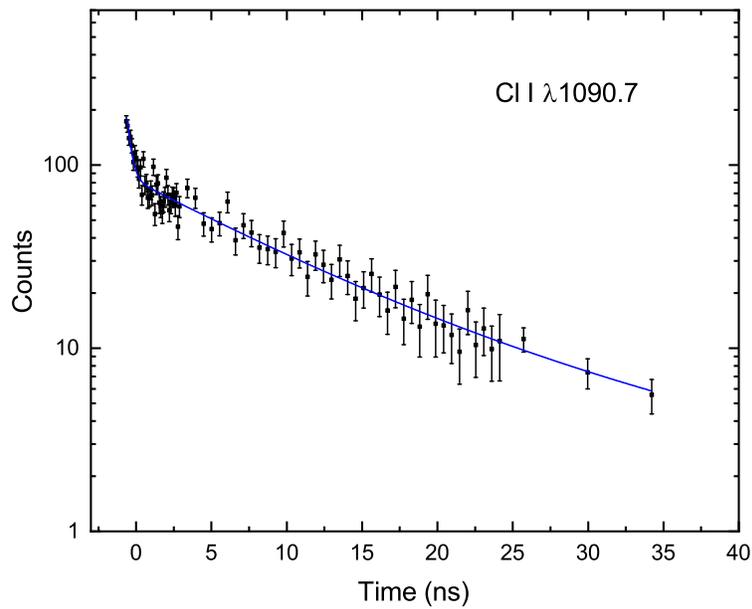}
\caption{Decay curve for the line at 1090.73 \AA\ for a beam energy of 170 keV utilizing a three-exponential fit as shown by the solid curve. At the shortest time, the decay from Cl~{\footnotesize V} is seen, and the portion between 22 and 35 ns  reveals cascades of about 30 ns. The step size furthest from the slit was 4 mm.}
\label{fig:f3}
\end{figure}

The situation was more complicated for the decay of the 3$s^2$3$p^4$5$s$ $^2P_{3/2}$ level from the line at 1090.73 \AA. The cascades from higher levels could impact our lifetime measurements by affecting the population of the level of interest. The results from \cite{oli13} suggest that decays from many upper states may populate this level. The common cascades involving  3$p^4$4$p$ $^2D_{3/2,5/2}^{\rm o}$ levels with a typical lifetime of 30 ns were observed in our decay curve. Evidence for other cascades was not apparent; those with lifetimes greater than 30 ns are beyond the capabilities of THIA. We applied the method of Arbitrarily Normalized Decay Curves (ANDC; \cite{cur71}) to remove the effect of cascades from the primary mean lifetime. Because the lifetimes of the possible cascades were significantly longer than the primary decay, the cascades played a minor role and affected the primary lifetime by less than 16$\%$. A semi$-$log plot for the 5$s$ $^2P_{3/2}$ decay curve is shown in Figure~\ref{fig:f3} using the $\lambda$1090.73 transition. 

A short-lived decay component from the fit to a transition in Cl~{\footnotesize V} at 545.11 \AA\ (3$s$3$p$ $^4P^{\rm o}_{5/2}$) was seen in  second order. We only saw the effect of Cl~{\footnotesize V} in the decay curve at an energy of 170 keV (see Figure~\ref{fig:f3}) because (1) measurements at higher energy produce higher charge states and (2) the slight difference in wavelength setting between the measurements at 130 and 170 keV could have included more intensity from the Cl~{\footnotesize V} line at 170 keV. \cite{sch05} found contaminating lines from highly-ionized chlorine (Cl~{\footnotesize  IV} and {\footnotesize V}) in their THIA data on Cl~{\footnotesize II}, too. The same situation was observed by \cite{bro18} in their work on P~{\footnotesize II}. Further measurements at 170 keV were produced on the Cl~{\footnotesize V} line at 545.11 \AA\ to establish its short lifetime.

\begin{figure}[ht]
\centering
\includegraphics[width=0.7\textwidth]{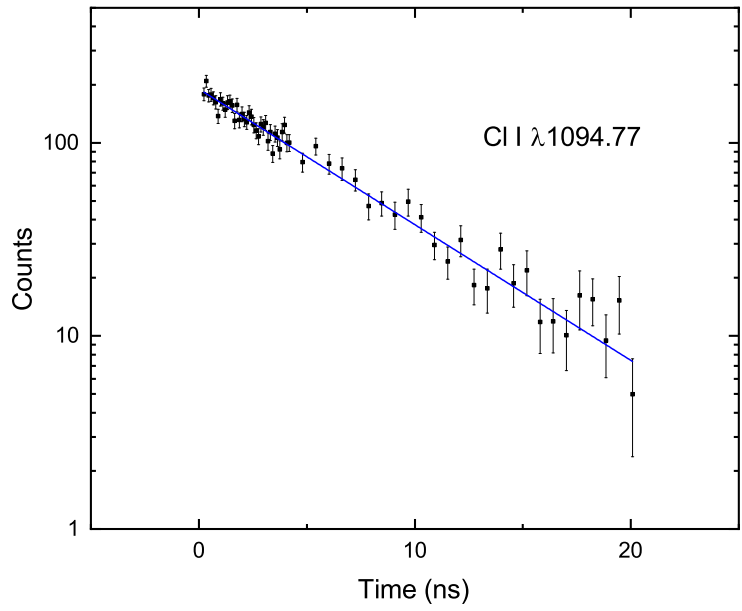}
\caption{Decay curve for the line at 1094.77\AA\ for a beam energy of 130 keV.}
\label{fig:f4}
\end{figure}

For the line at 1094.77 \AA, resulting from decay  of the 3$s^2$3$p^4$3$d$  $^2F_{5/2}$ level, the decay curve may be affected by contamination from adjacent lines. The possible contaminating intensity that may be coming from Cl~{\footnotesize I} lines at 1095.14 and 1095.66 \AA\ was studied by moving the grating position. One setting at 1094.53 \AA\ was far away from any blending, while a setting at 1095.43 \AA\ was where the most significant amounts of contamination were expected. The fits yielded indistinguishable differences in lifetimes, indicating there was no detectable contamination from other Cl~{\footnotesize I} lines. The best single-exponential fit is shown in Figure~\ref{fig:f4} for the $\lambda$1094.77 transition.

We attempted to measure branching ratios from the ratio of intensities for the pairs of lines, 1079.88 versus 1090.27 \AA\ and 1090.73 versus 1103.34 \AA.  However, our spectral measurements were not precise enough for any meaningful results because the blending and contamination from Cl~{\footnotesize  IV} and {\footnotesize V} seen in second order, and the resulting limited signal to noise during the attempt to remove the contamination, prevented us from making accurate branching fractions from the intensity ratios.  As a result, we adopted the branching ratios for $\lambda\lambda$1079.88, 1090.73 from the calculations of \cite{oli13} when inferring oscillator strengths from our experimental lifetimes.  The decays associated with the other two transitions at 1004.68 and 1094.77 \AA\ represent the dominant channel according to \cite{oli13}.

\section{Results and Discussion}

Our results for lifetime measurements, as well as comparison with earlier theoretical calculations, are shown in Table~\ref{table:t1}. The weighted average of the lifetimes reported in Table~\ref{table:t1} are obtained from experiments acquired at beam energies 130 and 170 keV. Our lifetime measurements are in very good agreement with theoretical calculations by \citet{sin06} and \citet{oli13}, considering our 2$\sigma$ uncertainties lead to differences of less than 15$\%$. These theoretical calculations are based on configuration interaction wavefunctions and relativistic effects. The earlier effort by \citet{ojh90} yielded much longer lifetimes due to the heavy mixing among Cl~{\footnotesize I} levels encountered in their less extensive calculations.

Our lifetime for the 3$s$3$p$ $^4P^{\rm o}_{5/2}$ level of  Cl~{\footnotesize V} is $0.27 \pm 0.01$ ns. Our value does not agree with theoretical results of approximately  0.11 ns obtained by several methods \citep{faw87,cho03,fro06}. Perhaps our measurements were affected by cascades, yielding a longer lifetime for 3$s$3$p$ $^4P^{\rm o}_{5/2}$. In any case, its approximate value helps us to extract the primary lifetime of interest.

Finally, we compare results for oscillator strengths derived from experimental lifetimes with those published previously, as shown in Table~\ref{table:t2}.~Generally, our experimental values for oscillator strengths are in very good agreement with those of \cite{son06}, \cite{sin06}, and \cite{oli13}, again with differences less than about 15$\%$. The correspondence is less satisfactory with the calculations of \cite{ojh90} and \cite{bie94}.

A reason for the poorer agreement with the results from \cite{ojh90} and \cite{bie94} may be the smaller scale of the computations compared with what is possible with today's servers. \cite{ojh90} predicted oscillator strengths with smaller values than other $f$-values shown in Table~\ref{table:t2}. The less extensive set of calculations leads to fewer terms in the eigenfunction expansions, causing the amount of mixing between the fine structure levels in the 3$p^4 (^3P)$ 3$d$ and 3$p^4 (^3P)$ 5$s$ configurations to be underestimated. This is likely the cause for the significant differences between the length and velocity results in their calculations. The same explanation may apply to the results from \cite{bie94} because the level of mixing could produce both smaller and larger $f$-values. The $\lambda$1094.77 transition is more complicated  theoretically because the 3$p^4$($^3P$) 3$d$ configuration mixes with 3$p^4$($^3P$) 5$s$ to produce 22 even parity levels. \cite{oli07} changed the designation of the heavily mixed level ($J=5/2$) to be ($^3P$)3$d$ $^2F_{5/2}$ and that improved their theoretical calculations. Since \cite{bie94} kept the term classification given in the NIST tables \citep{kra08}, some of the difference between their $f$-value for the line at 1094.77 \AA\ and ours may arise from the classification scheme used. 

While our results generally agree with the $f$-values given by \cite{son06}, that is not the case for $\lambda$1090.73. Sonnentrucker (2019, private communication) noted that this portion of the spectrum contains substantial amounts of  blending from a combination of stellar features and the wing of a strong H$_2$ line. The line blends likely affected their placement of the stellar continuum when measuring this relatively weak Cl~{\footnotesize I} line.

\section{Conclusions}
We presented experimental lifetimes and the corresponding $f$-values based on beam-foil spectroscopic techniques that are needed in studies of interstellar Cl~{\footnotesize I}. Our results agree very well with the most recent theoretical predictions  of \citet{sin06} and \citet{oli13}, and are generally consistent with the $f$-values obtained from astronomical spectra \citep{son06}. The resulting $f$-values are in quite good agreement with those given by  \cite{sin06}, \cite{son06}, and \cite{,oli13},  except the $f$-value for the line at 1090.73 \AA\ that does not agree with \cite{son06}. In particular, these transitions provide important information on the structure of diffuse gas because Cl~{\footnotesize I} is a good tracer of molecular components. On the other hand, our $f$-values differ substantially from the theoretical calculations presented by \cite{ojh90} and \cite{bie94}. The results of our lifetime measurements and oscillator strengths are the only available experimental measurements for these levels to date. Since our results agree so well with the latest theoretical efforts, future analyses can incorporate these additional Cl~{\footnotesize I} transitions for improved self-consistency.

\acknowledgments
We thank Paule Sonnentrucker for useful comments and Robert Snuggs for technical assistance.

\bibliographystyle{plainnat}

\begin{thebibliography}{}
\bibitem[Bi\'emont et al.(1994)]{bie94} Bi\'emont, E., Gebarowski, R., \& Zeippen, C. ~J. 1994,  A\&A, 287, 290
\bibitem[Brown et al.(2018)]{bro18} Brown, M. ~S., Alkhayat, R. ~B., Irving, R. ~E., et al. 2018, ApJ, 868, 42
\bibitem[Choudhury et al.(2003)]{cho03} Choudhury, K.~B, Deb, N.~C., Roy, K., \& Msezane A.~Z. 2003, Eur Phys JD, 27, 103
\bibitem[Curtis et al.(1971)]{cur71} Curtis, L.~J., Berry, H.~G., \& Bromander, J.\ 1971, PhLA, 34, 169
\bibitem[Dalgarno et al.(1974)]{dal74}Dalgarno, A., De Jong, T., Oppenheimer, M., \& Black, J.~H. 1974, ApJ, 192, L37
\bibitem[Fawcett(1987)]{faw87} Fawcett, B. ~C. 1987, ADNDT, 37, 411
\bibitem[Federman(1986)]{fed86} Federman, S.~R. 1986, ApJ, 309, 306
\bibitem[Federman et al.(1992)]{fed92} Federman, S.~R., Beideck, D.~J., Schectman, R.~M., \& York, D.~G. 1992, ApJ, 401, 367
\bibitem[Froese Fischer et al.(2006)]{fro06} Froese Fischer, C., Tachiev, G., \& Irimia, A. 2006, ADNDT, 92, 607
\bibitem[Haar et al.(1993)]{haa93}Haar, R.~R., Beideck, D.~J., Curtis, L.~J., et al. 1993, NIMPB, 79, 746
\bibitem[Jura(1974)]{jur74} Jura,  M. 1974, ApJ, 190, L33
\bibitem[Jura \& York(1978)]{jur78} Jura,  M., \& York, D.G. 1978, ApJ, 219, 861
\bibitem[Kramida et al.(2018)]{kra08} Kramida, A., Ralchenko, Yu., Reader, J., and NIST ASD Team. 2018. NIST Atomic Spectra Database (ver. 5.6.1). \url{https://physics.nist.gov/asd3}
\bibitem[Moomey et al.(2012)]{moo12} Moomey, D., Federman, S.~R., \& Sheffer, Y. 2012, ApJ, 744, 174
\bibitem[Morton(2003)]{mor03} Morton, D.~C.\ 2003,\ ApJS, 149, 205
\bibitem[Neufeld \& Wolfire(2009)]{neu09} Neufeld,  D.~A., \& Wolfire, M.~G. 2009, ApJ, 706, 1594 
\bibitem[Ojha \& Hibbert(1990)]{ojh90} Ojha, P. C., \& Hibbert, A. 1990, Phys Scr, 42, 424
\bibitem[Oliver \& Hibbert(2007)]{oli07} Oliver, P., \& Hibbert, A. 2007, JPhB, 40, 2847
\bibitem[Oliver \& Hibbert(2008)]{oli08} Oliver, P., \& Hibbert, A. 2008, J. Phys: Conf. Ser. 130 012016
\bibitem[Oliver \& Hibbert(2013)]{oli13} Oliver, P., \& Hibbert, A. 2013, ADNDT, 99, 459
\bibitem[Schectman et al.(2000)]{sch00} Schectman, R.M., Cheng, S., Curtis, L.J., et al. 2000, ApJ, 542, 400
\bibitem[Schectman et al.(1993)]{sch93} Schectman, R.M., Federman, S.~R., Beideck, D.~J., \& Ellis, D. ~G. 1993, ApJ, 406, 735
\bibitem[Schectman et al.(2005)]{sch05} Schectman, R.M., Federman, S.~R., Brown, S., et al. 2005, ApJ, 621, 1159
\bibitem[Singh et al.(2006)]{sin06} Singh,~N., Jha,~A.~K.~S., \& Mohan,~A. 2006, Eur Phys JD, 38, 285
\bibitem[Sonnentrucker et al.(2002)]{son02} Sonnentrucker, P., Friedman, S.~D., Welty, D.~E., et al. 2002, ApJ, 576, 241
\bibitem[Sonnentrucker et al.(2006)]{son06} Sonnentrucker, P., Friedman, S.~D., \& York, D.~G. 2006, ApJ, 650, L115



\end{thebibliography}


\begin{deluxetable}{cccccccccc} 
\tablecolumns{6}
\tablewidth{0pt}
\tabletypesize{\footnotesize}
\tablecaption{Cl~{\footnotesize  I} Lifetimes}
\tablehead{\colhead{$J_u$} & \colhead{$\lambda_{ul}$}& \multicolumn{4}{c}{$\tau$ (ns)} \\
\cline{3-6} \\
\colhead{} & \colhead{(\AA)} & \colhead{Present} & \colhead{OH\tablenotemark{a}} & \colhead{SJM\tablenotemark{b}}&\colhead{O\tablenotemark{c}}  
}
\startdata
1/2 & 1004.68 & $1.04 \pm 0.08$ & $\ldots$  & $1.06$ & $1.11$  \\
1/2 & 1079.88& $5.83 \pm0.36$ & $21.26$ & $\ldots$  & $6.26$ \\
3/2 & 1090.73 & $8.98 \pm0.72$ & $17.77$  & $\ldots$  & $8.07$ \\
5/2 &1094.77 & $6.28 \pm 0.18$ & $29.23$  & $6.21$  & $7.50$ \\
\enddata
\tablenotetext{a}{Ojha \& Hibbert (1990) $-$ configuration interaction using CIV3.}
\tablenotetext{b}{Singh et al. (2006) $-$ configuration interaction using CIV3.}
\tablenotetext{c}{Oliver \& Hibbert (2013) $-$ configuration interaction using CIV3.}
\label{table:t1}
\end{deluxetable}

\begin{deluxetable}{cccccccccccccc}
\tablecolumns{8}
\tablewidth{0pt} 
\tabletypesize{\footnotesize}
\tablecaption{Cl~{\footnotesize I} Oscillator Strengths}
\tablehead{\colhead{$\lambda_{ul}$ (\AA)} & \colhead{$J_l$} & \colhead{$J_u$} & \multicolumn{7}{c}{$f$-value ($\times 10^{-2}$)} \\
\cline{4-10} \\
\colhead{} & \colhead{} &  \colhead{} & \colhead{Present} & \colhead{OH\tablenotemark{a}} & \colhead{BGZ\tablenotemark{b}}&\colhead{SFY\tablenotemark{c}}& \colhead{SJM\tablenotemark{d}}&\colhead{O\tablenotemark{e}} 
}
\startdata
1004.68 & 3/2 & 1/2 & $4.73 \pm 0.36$ & $\ldots$ & $\ldots$ & $5.14^{+0.58}_{-0.51}$ & 4.69\tablenotemark{f} & 4.43\tablenotemark{f} \\
{       }& $\ldots$ & $\ldots$ & $\ldots$ & $\ldots$ & $\ldots$ &$\ldots$ & 3.99\tablenotemark{g} & 4.25\tablenotemark{g} \\
1079.88 & 3/2 & 1/2 & $0.47 \pm 0.03$ & 0.065\tablenotemark{f} & 0.70\tablenotemark{f} &  $0.56^{+0.13}_{-0.12}$ & $\ldots$ & 0.44\tablenotemark{f} \\
{       } & $\ldots$ & $\ldots$  & $\ldots$  & 0.13\tablenotemark{g} & 0.69\tablenotemark{g} &$\ldots$ & $\ldots$& 0.40\tablenotemark{g} \\
1090.73 & 3/2 & 3/2 & $0.95 \pm 0.08$  & 0.25\tablenotemark{f} & 0.68\tablenotemark{f} & $0.28^{+0.04}_{-0.04}$ & $\ldots$ & 1.07\tablenotemark{f} \\
{       }& $\ldots$ &  $\ldots$ & $\ldots$ & 0.46\tablenotemark{g} & 0.61\tablenotemark{g} & $\ldots$ &$\ldots$ & 0.97\tablenotemark{g}  \\
1094.77 & 3/2 & 5/2 & $3.85 \pm 0.11$ & 0.11\tablenotemark{f} &  1.66\tablenotemark{f} & $3.96^{+0.42}_{-0.38}$ &  3.72\tablenotemark{f} &  3.22\tablenotemark{f} \\
{       } & $\ldots$  & $\ldots$ & $\ldots$ & 0.045\tablenotemark{g} & 1.52\tablenotemark{g} & $\ldots$ & 3.94\tablenotemark{g} & 3.04\tablenotemark{g} & \\
\
\enddata
\tablenotetext{a}{Ojha \& Hibbert (1990).}
\tablenotetext{b}{Bi\'emont et al. (1994) $-$ configuration interaction using SUPERSTRUCTURE.}
\tablenotetext{c}{Sonnentrucker et al. (2006) $-$ astronomical observations.} 
\tablenotetext{d}{Singh et al. (2006).} 
\tablenotetext{e}{Oliver \& Hibbert (2013).}
\tablenotetext{f}{Based on length formalism.}
\tablenotetext{g}{Based on velocity formalism.}
\label{table:t2}
\end{deluxetable}
\clearpage

\clearpage

\end{document}